\documentclass[dvips]{article}

\usepackage{icrc2011}

\title{Measurement of Cosmic Ray $\bar{p}/p$ flux ratio at TeV energies with ARGO-YBJ}

\newcommand{\etal}{\MakeLowercase{\textit{et al. }}} 
\shorttitle{author \etal paper short title}

\authors{G. Di Sciascio$^{1}$ and R. Iuppa$^{1,2}$ for the ARGO-YBJ collaboration}
\afiliations{$^1$INFN, Sezione Roma Tor Vergata, Via della Ricerca Scientifica 1, Roma, Italy\\ $^2$ Dipartimento di Fisica, Universit\'a Roma Tor Vergata, via della Ricerca Scientifica 1, Roma, Italy}
\email{disciascio@roma2.infn.it, iuppa@roma2.infn.it}

\abstract{Cosmic ray antiprotons provide an important probe for the study of cosmic-ray propagation in the interstellar space and to investigate the existence of Galactic dark matter.  The ARGO-YBJ experiment, located at the Yangbajing Cosmic Ray Laboratory (Tibet, P.R. China, 4300 m a.s.l., 606 g/cm$^2$ ), is the only experiment exploiting the full coverage approach at very high altitude presently at work. The ARGO-YBJ experiment is particularly effective in measuring the cosmic ray antimatter content via the observation of the cosmic rays Moon shadowing effect.
Based on all the data recorded during the period from July 2006 through November 2009 and a full Monte Carlo simulation, we searched for the existence of the shadow produced by antiprotons at the few-TeV energy region. No evidence of the existence of antiprotons was found in this energy region. Upper limits to the antip/p flux ratio are set to 5\% at a median energy of 2 TeV and 6\% at 5 TeV with a confidence level of 90\%. In the few-TeV energy range this result is the lowest available.}

\begin{document}
\maketitle

\section{Introduction}

Very High Energy Cosmic Ray (VHE CR) antiprotons are an essential diagnostic tool to approach the solution of several big topics of cosmology, astrophysics and particle physics, besides for studying fundamental properties of the CR sources and propagation medium \cite{stecker02}. The enigma of the matter/anti-matter asymmetry in the local Universe, namely the existence of antimatter regions, the signatures of physics beyond the standard model of particles and fields, as well as the determination of the essential features of CR propagation in the insterstellar medium, are a few research topics which would greatly benefit from the detection of VHE antiprotons \cite{stecker84,klopov00,strong07,pamelap10,cirelli,donato09,golden}. 

Cosmic rays are hampered by the Moon, therefore a deficit of CRs in its direction is expected (the so-called \emph{Moon shadow}). Moreover, the Earth-Moon system acts as a magnetic spectrometer. In fact, due to the Geomagnetic field (GMF) the Moon shadow shifts westward by an amount depending on the primary CR energy. The paths of primary antiprotons are therefore deflected in the opposite sense in their way to the Earth. This effect allows, in principle, the search for antiparticles in the opposite direction of the observed Moon shadow.

If the energy is low enough and the angular resolution is good we can
distinguish the two shadows, one shifted towards West due to the protons and the other shifted towards East due to the antiprotons. 
At high energies ($\geq$ 10 TeV) the
magnetic deflection is too small compared to the angular
resolution and the two shadows cannot be disentangled. At low energies
($\leq$500 GeV) the shadows are much deflected but washed out by
the poor angular resolution, thus the sensitivity is limited.
Therefore, there is an optimal energy window for the measurement
of the antiproton abundance.

The ARGO-YBJ experiment is especially recommended for the measurement of the CR antimatter content via the observation of the Galactic CR shadowing effect due to: (1) good angular resolution and pointing accuracy and their long term stability; (2) low energy threshold; (3) real sensitivity to the GMF.
Indeed, the low energy threshold of the detector allows the observation of
the shadowing effect with a sensitivity of about 9 standard deviations (s.d.) per month at TeV energy.

In this paper we report the measurement of the CR $\overline{p}/p$
flux ratio in the TeV energy region with all the data collected during the period from July 2006 to November 2009.

\section{The ARGO-YBJ experiment}

The detector is composed of a central carpet large $\sim$74$\times$
78 m$^2$, made of a single layer of Resistive Plate Chambers
(RPCs) with $\sim$93$\%$ of active area, enclosed by a guard ring
partially ($\sim$20$\%$) instrumented up to $\sim$100$\times$110
m$^2$. The apparatus has modular structure, the basic data
acquisition element being a cluster (5.7$\times$7.6 m$^2$),
made of 12 RPCs (2.8$\times$1.25 m$^2$ each). Each chamber is
read by 80 external strips of 6.75$\times$61.8 cm$^2$ (the spatial pixels),
logically organized in 10 independent pads of 55.6$\times$61.8
cm$^2$ which represent the time pixels of the detector. 
The read-out of 18360 pads and 146880 strips are the experimental output of the detector \cite{aielli06}.
The central carpet contains 130 clusters (hereafter, ARGO-130) and the
full detector is composed of 153 clusters for a total active
surface of $\sim$6700 m$^2$. 

The whole system, in smooth data taking since July 2006 firstly with ARGO-130, is in stable data taking with the full apparatus of 153 clusters since November 2007 with the trigger condition N$_{trig}$ = 20 and a duty cycle $\geq$85\%. The trigger rate is $\sim$3.5 kHz with a dead time of 4$\%$.

The analysis reported in this paper refers to events collected
after the following selections: (1) more than 20 strips on the ARGO-130 carpet; (2) zenith angle of the shower arrival direction less than 50$^{\circ}$; (3) reconstructed core position inside a 150$\times$150 m$^2$ area centered on the detector. 
According to the simulation, the median energy of the selected protons is E$_{50}\approx$1.8 TeV (mode energy $\approx$0.7 TeV).

\section{Monte Carlo simulation}

A detailed Monte Carlo (MC) simulation of the CR propagation in the Earth-Moon system \cite{mcmoon-icrc09} has been developed in order to estimate the expected antiproton flux in the side opposite to the CR Moon shadow.
The simulation is based on the real data acquisition time. The Moon position has been computed at fixed times, starting from July 2006 up to November 2009. Such instants are distant 30 seconds each other. For each time, after checking that the data acquisition was effectively running and the Moon was in the field of view, primaries were generated with arrival directions sampled within the Moon disc basing on the effective exposure time.

After accounting for the arrival direction correction due to the magnetic bending effect, the air shower development in the atmosphere has been generated by the CORSIKA v. 6.500 code with QGSJET/GHEISHA models \cite{corsika}. 
CR spectra have been simulated in the energy range from 10 GeV to 1 PeV following the results given in \cite{horandel}. About 10$^8$ showers have been sampled in the zenith angle interval 0-60 degrees. 
The experimental conditions (trigger logic, time resolution, electronic noises, etc.) have been reproduced by a GEANT4-based code \cite{geant4}. 

\section{Data analysis}

For the analysis of the shadowing effect, the signal is collected within a 10$^{\circ}\times$10$^{\circ}$ sky region centered on the Moon position. We used celestial coordinates (right ascension and declination, hereafter R.A. and DEC.) to produce the \emph{event} and \emph{background} sky maps, with 0.1$^{\circ}\times$0.1$^{\circ}$ bin size.
Finally, after a smoothing procedure, the \emph{significance} map, used to estimate the statistical significance of the observation, is built.

\begin{figure}[!htpb]
  \centering
  \includegraphics[width=0.48\textwidth]{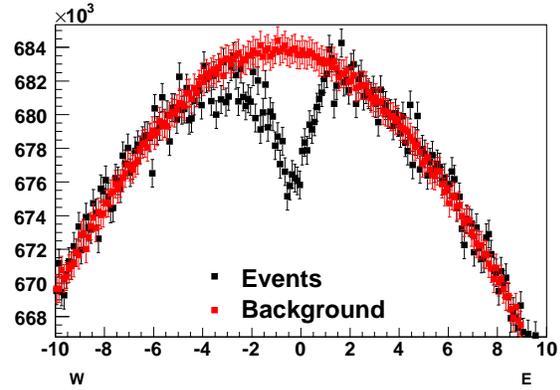}
  \caption{Deficit of CRs around the Moon position projected along the R.A. direction. Showers with N$>$60 recorded from July 2006 until November 2009 are shown.}
  \label{fig:moon-evbkg}
 \end{figure}

As can be noticed from Fig. \ref{fig:moon-evbkg}, the Moon shadow turns out to be a lack in the smooth CR signal observed by ARGO-YBJ, even without subtracting the background contribution. The background events are not uniformly distributed around the Moon, because of the non-uniform exposure of the map bins to CR radiation. 
 
The background has been estimated with two different approaches, the \emph{time swapping} and the \emph{equi-zenith angle} methods, as described in \cite{aielli10}, in order to investigate possible systematic uncertainties in the background calculation. They give results consistent with each other within 1 s.d..

 \begin{figure}[!htbp]
  \centering  \includegraphics[width=0.48\textwidth]{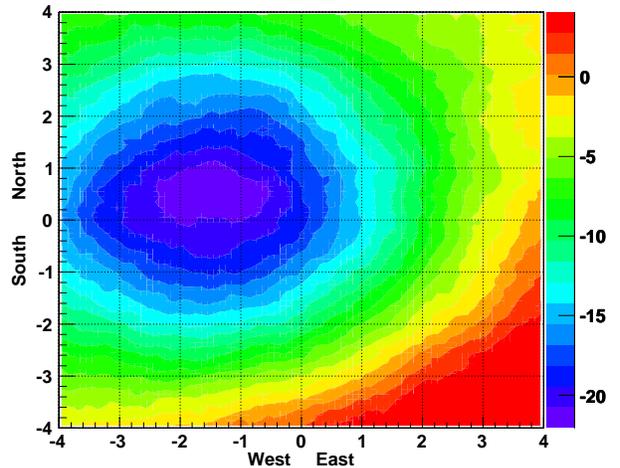}
  \caption{Moon shadow significance map. It collects all the events detected by ARGO-YBJ from July 2006 until November 2009. The event multiplicity is 20$\leq $N$<$40 and zenith angle $\theta<50^{\circ}$. The color scale gives the statistical significance.}
  \label{fig:moon1}
 \end{figure}
%

A significance map of the Moon region is shown in Fig. \ref{fig:moon1}. It contains all the events belonging to the lowest multiplicity bin investigated (20$\leq$N$<$40), collected by ARGO-YBJ during the period July 2006 - November 2009 (about 3200 hours on-source in total). The significance of the maximum deficit is about 22 s.d..
The observed westward displacement of the Moon shadow by about 1.5$^{\circ}$ allows to appreciate the sensitivity of the ARGO-YBJ experiment to the GMF.
\emph{In such a map a potential antiproton signal is expected eastward within 1.5$^{\circ}$ from the actual Moon position} (i.e., within 3$^{\circ}$ from the observed Moon shadow). 
The median energy of the selected events is E$_{50}\approx$750 GeV (mode energy $\approx$ 550 GeV) for proton-induced showers. The corresponding angular resolution is $\sim$1.9$^{\circ}$.
However, the large displacement of the shadow is only one ingredient of this analysis, the other being the angular resolution which is not small enough in this multiplicity range. Indeed, as can be seen from Fig. \ref{fig:moon1}, the matter shadow extends to the antimatter side with a significance of about 10 s.d., thus limiting the sensitivity to the antiproton abundance measurement.

We stress that this is the first time that an EAS array is able to detect the Moon shadow so shifted, observing the signal due to sub-TeV primary CRs.

%
 \begin{figure}[!t]
  \centering
  \includegraphics[width=0.48\textwidth]{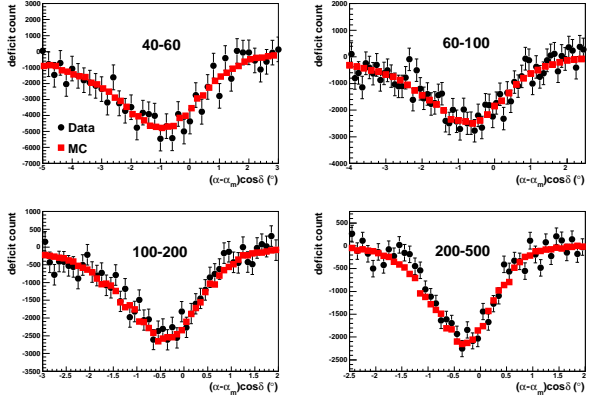}
  \caption{Deficit counts measured around the Moon projected along
the East-West axis for different multiplicity bins (black circles) compared to MC expectations (red squares).}
  \label{fig:proj-ew}
 \end{figure}

\section{Results and discussion}

The optimal energy windows for the measurement of the antiproton abundance in CRs are given by the following multiplicity ranges: 40$\leq$N$<$100 and N$\geq$100.
In the former bin the statistical significance of Moon shadow observation is 34 s.d., the measured angular resolution is $\sim$1$^{\circ}$, the proton median energy is 1.4 TeV and the number of deficit events about 183000.
The Moon shadow is shifted by 0.82$^{\circ}$ westward from its actual position.
In the latter multiplicity bin the significance is 55 s.d., the measured angular resolution $\sim$0.6$^{\circ}$, the proton median energy is 5 TeV and the number of deficit events about 46500.
The Moon shadow is shifted by 0.30$^{\circ}$ westward.

The chance of unfolding all CR spectral contributions relies on MC simulations, as well as the search for antiprotons demands to properly reproduce the Moon shadow signal.
The deficit counts observed around the Moon projected on the
East-West axis are shown in Fig. \ref{fig:proj-ew} for 4 multiplicity bands compared to MC expectations.
We used the events contained in an angular band parallel to the East-West axis and centered on the observed Moon position. The widths of these bands are chosen on the basis of the MC simulation so that the shadow deficit is maximized. They turn out to be proportional to the N$_{strip}$-dependent angular resolution: $\pm$2.9$^{\circ}$ in
40$\leq N_{strip} <$ 60, $\pm$2.6$^{\circ}$ in 60$\leq N_{strip}
<$ 100, $\pm$2.1$^{\circ}$ in 100$\leq N_{strip} <$ 200,
$\pm$1.6$^{\circ}$ in 200$\leq N_{strip} <$ 500. 
As an expected effect of the GMF, the profile of the shadow is
broadened and the peak positions shifted westward as the
multiplicity (i.e., the cosmic ray primary energy) decreases. The
data are in good agreement with the expectations from the MC
simulation. 

In order to evaluate the $\bar{p}/p$ ratio, the projections of the Moon shadow along the R.A. direction have been considered. The GMF shifts westward the dip of the signal from positively charged primaries. Searching antiprotons means looking for excesses in the eastern part of the R.A. projection, i.e. trying to fit the Moon shape expected from combining CR and antimatter to the shape obtained from the experimental data. It is worth noticing that for the antiprotons we assume the same energy spectrum of the protons.
Of course, whichever matter-antiprotons combination is obtained, the total amount of triggered events must not be changed, so that the fitting procedure consists in transferring MC events from the CRs to the antiproton shadow and comparing the result with the data.

To make such a comparison, we firstly adopted the following method. We obtained two kinds of Moon shadow, cast by all CRs and protons, respectively. After projecting them along the R.A. direction, we used a superposition of several Gaussian functions to describe the deficit event distribution in each shadow. Four Gaussian functions were found to be adequate for fitting both distributions within 5$^{\circ}$ from the Moon disc center. Let us name $\theta$ the angular distance from the Moon disc center and $f_m(\theta)$ the Gaussian function superposition describing the CR shadow. Let $\mathcal{F}_{p}(\theta)$ be the proton shadow, obtained by imposing a given power law spectrum. The observed Moon shadow should be expressed by the following function:
%
  \setlength{\arraycolsep}{0.0em}
  \begin{eqnarray*}
f_{MOON}(\theta)=(1-r)\,f_{m}(\theta)+r\,\mathcal{F}_{\overline{p}}(\theta)
\\
=(1-r)\,f_{m}(\theta)+r\,\mathcal{F}_{p}(-\theta)
  \end{eqnarray*}
  \setlength{\arraycolsep}{5pt}
%
($0\leq r<1$) where the first term represents the deficit in CRs and the second term represents the deficit in antiprotons. This function must be fitted to the data to obtain the best value of $r$. 

We also applied a second method to determine the antiproton content in the cosmic radiation. Without introducing functions to parameterize the expectations, we directly compared the MC signal with the data. We performed a Maximum Likelihood fit using the $\bar{p}$ content as a free parameter with the following procedure:
\begin{enumerate}
 \item the Moon shadow R.A. projection has been drawn both for data and MC.
 \item the MC Moon shadow has been split into a ``matter'' part
 \emph{plus} an ``antiproton'' part, again so that the
 \emph{total amount of triggered events remains unchanged}:
  \setlength{\arraycolsep}{0.0em}
  \begin{eqnarray*}
\Phi_{MC}(mat) \longrightarrow \Phi_{MC}(r;mat+\bar{p})=
\\
= (1-r)\Phi_{MC}(mat)+\Phi_{MC}(\bar{p})
  \end{eqnarray*}
  \setlength{\arraycolsep}{5pt}
 \item for each antiproton to matter ratio, the expected Moon shadow
R.A. projection $\Phi_{MC}(r;mat+\bar{p})$ is compared with the
experimental one via the calculation of the likelihood function:
%
$$ log\mathcal L (r)=\sum_{i=1}^B N_i ln[E_i(r)]-E_i(r)-ln(N_i!)$$
%
where N$_i$ is the number of experimental events included
within the $i$-th bin, while $E_i(r)$ is the number of
events expected within the same bin, which is calculated by adding the contribution expected from MC ($\Phi_{MC}(r;mat+\bar{p})$) to the \emph{measured} background.
\end{enumerate}

Both methods described above give results consistent within 10\%. The $r$ parameter which best fits the expectations to the data turns out to be always negative, i.e. it assumes non-physical values throughout the whole energy range investigated. 
With a direct comparison of the R.A. projections, the $r$-values which maximize the likelihood are: -0.076$\pm$0.040 and -0.144$\pm$0.085 for 40$\leq$N$<$100 and N$\geq$100, respectively. The corresponding upper limits with 90\% confidence level (c.l.), according to the unified Feldman \& Cousins approach \cite{feldman}, are 0.034 and 0.041, respectively.
Since the anti-shadow was assumed to be the mirror image of the proton shadow, we assume for the antiprotons the same median energy. 
The $\bar{p}/p$ ratio is $\Phi(\bar{p})/\Phi(p)$ = 1/$f_p\cdot$ $\Phi(\bar{p})/\Phi(matter)$, therefore, being the assumed proton fraction $f_p$=73\% for 40$\leq$N$<$100 and $f_p$=71\% for N$\geq$100 \cite{horandel}, we obtain the following upper limits at 90\% c.l.: 0.05 for 40$\leq$N$<$100 and 0.06 for N$\geq$100.
Notice that the two values are similar, in spite of the different multiplicity interval. It is a consequence of the combination of the two opposing effects of the angular resolution and of the geomagnetic deviation.

\begin{figure}
  \centering
  \includegraphics[width=0.5\textwidth]{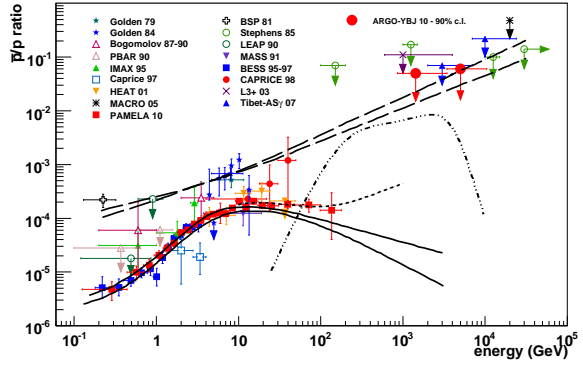}
  \caption{The $\bar{p}/p$ flux ratio obtained with the
ARGO-YBJ experiment compared with all the available measurements and some theoretical predictions (see text). }
  \label{AntiProton}
\end{figure}

In Fig. \ref{AntiProton} the ARGO-YBJ results are shown with all the available measurements. The solid curves refer to a theoretical predictions for a pure secondary production of antiprotons during the CR propagation in the Galaxy by Donato et al. \cite{donato09}. The curves was obtained using the appropriate solar modulation parameter for the PAMELA data taking period \cite{pamelap10}.
The long-dashed lines refer to a model of primary $\bar{p}$ production by antistars \cite{golden}. The rigidity-dependent confinement of CRs in the Galaxy is assumed $\propto$ R$^{-\delta}$, and the two curves represent the cases of $\delta$ = 0.6, 0.7. The dot-dashed line refers to the contribution of $\bar{p}$ from the annihilation of a heavy dark matter particle \cite{cirelli}.
The short-dashed line shows the calculation by Blasi and Serpico \cite{blasi2} for secondary antiprotons including an additional $\bar{p}$ component produced and accelerated at CR sources.
%

\section{Conclusions}

The ARGO-YBJ experiment is observing the Moon shadow with high statistical significance at an energy threshold of a few hundred GeV. Using all data collected until November 2009, we set two upper limits on the $\bar{p}/p$ flux ratio: 5\% at an energy of 1.4 TeV and 6\% at 5 TeV with a confidence level of 90\%. 
In the few-TeV range the ARGO-YBJ results are the lowest available, useful to constrain models for antiproton production in antimatter domains.


\end{document}